\documentclass[
aps,nofootinbib,showpacs,showkeys,preprint
,tightenlines ] {revtex4}

\usepackage{epsf,epsfig,subfigure,graphicx,amsmath,amssymb}
\usepackage{color}

\newcommand{\dis}[1]{\begin{equation}\begin{split}#1\end{split}\end{equation}}
\newcommand{\be}{\begin{equation}}
\newcommand{\ee}{\end{equation}}

\newcommand{\eq}[1]{Eq.~(\ref{#1})}

\def\bea{\begin{eqnarray}}
\def\eea{\end{eqnarray}}

\begin{document}

\title{\Large\bf 
Relaxing the Landau-pole constraint in the NMSSM 
\\with the Abelian gauge symmetries 
}

\author{Bumseok Kyae$^{(a)}$\footnote{email: bkyae@pusan.ac.kr}
and Chang Sub Shin$^{(b)}$\footnote{email: csshin@apctp.org}
}
\affiliation{$^{(a)}$
Department of Physics, Pusan National University, Busan 609-735, Korea
\\
$^{(b)}$ Asia Pacific Center for Theoretical Physics, Pohang, Gyeongbuk 790-784, Korea
}

\begin{abstract}

In order to relax the Landau pole constraint on 
``$\lambda$'', which is a coupling constant between a singlet $S$ and the MSSM Higgs, $\lambda Sh_uh_d$ 
in the next-to-MSSM (NMSSM), and also maintain the gauge coupling unification,   
we consider perturbative U(1) gauge extensions of the NMSSM.  
For relatively strong U(1) gauge interactions down to low energies, we assign U(1) charges only to the Higgs and the third family of the chiral matter among the MSSM superfields. 
In the U(1)$_Z$ [U(1)$_Z\times$U(1)$_X$] extension, 
the low energy value of $\lambda$ can be lifted up to $0.85-0.95$ [$0.9-1.0$], depending on the employed charge normalizations, when $\lambda$ and the new gauge couplings are required not to blow up below the $10^{16}$ GeV energy scale. 
Introduction of extra vector-like superfields can induce the desired Yukawa couplings for the first two families of the chiral matter. 
We also discuss various phenomenological constraints 
associated with extra U(1) breaking.

\end{abstract}

\pacs{14.80.Da, 12.60.Fr, 12.60.Jv}

\keywords{Higgs mass, NMSSM, Landau-pole problem}
\maketitle


\section{Introduction}

The smallness of the Higgs mass and the gauge coupling unification are the two salient features noted in the minimal supersymmetric standard model (MSSM) \cite{MSSM}.  
Since the Higgs quartic coupling in the tree-level potential is given by the small gauge couplings in the MSSM, the relatively light Higgs is favored.  
By supersymmetry (SUSY) the small Higgs mass can be protected up to the fundamental scale. As a result, the standard model (SM) can be naturally embedded in a grand unified theory (GUT) at a very high energy scale.  
Indeed, the gauge coupling unification inferred from the renormalization group (RG) runnings of the three SM gauge couplings in the MSSM might imply the presence of such a unified theory at the GUT scale ($\approx 2.2\times 10^{16}$ GeV).  

Recently, CMS and ATLAS reported the observations of the SM Higgs-like signals at 126 GeV invariant mass \cite{CMS,ATLAS}. In fact, however, 126 GeV is too heavy as the Higgs mass in the MSSM: 
the Higgs mass at the tree level in the MSSM is predicted to be lighter even than the $Z$ boson mass, and so excessively large radiative correction to it for explaining the 126 GeV Higgs mass is indispensable.  However, it could result in a fine-tuning among the soft parameters \cite{MSSM}. 
In order to avoid such a fine-tuning, an extension of the MSSM would be necessary  
such that the tree level Hiss mass \cite{NMSSM,nmssm2,nmssm3,PQNMSSM,PQNMSSM2,King,U1extension,U1MSSM} or the radiative correction to it is enhanced \cite{KP,extraMatt}.

In the next-to-MSSM (NMSSM), the ``$\mu$ term'' in the MSSM superpotential is promoted to the trilinear term  \cite{NMSSM},  
\dis{ \label{Shh}
\lambda Sh_uh_d
}
by introducing a singlet $S$ and a dimensionless parameter $\lambda$. A vacuum expectation value (VEV) of $S$ can reproduce the MSSM $\mu$ term. Such a trilinear term in the superpotential yields the quartic term in the Higgs potential. It adds a sizable correction to the MSSM Higgs mass at the tree level: 
\dis{ \label{higgsmass}
m_h^2= m_Z^2{\rm cos}^22\beta + \lambda^2v_h^2{\rm sin}^22\beta +\Delta m_h^2 . 
}
Here, the first and the last terms indicate the tree level Higgs mass in the MSSM and radiative corrections, respectively. $v_h$ ($\equiv\sqrt{v_u^2+v_d^2}=174$ GeV) denotes the Higgs VEV.
The second term in Eq.~(\ref{higgsmass}) originates from Eq.~(\ref{Shh}) of the NMSSM, 
which is very helpful for raising the Higgs mass up to 126 GeV without a serious fine-tuning among the SUSY breaking soft mass parameters and the $Z$-boson mass, if $\lambda$ is of order unity. 

The RG analysis shows that the size of $\lambda$ monotonically increases with energy, and so it eventually meets a Landau pole (LP) at a higher energy scale, if $\lambda$ is too large at low energy. 
It is known that $\lambda$ in the NMSSM should be smaller than $0.7$ at the electroweak scale for $\lambda$ not to blow up below the GUT scale (``LP constraint'') \cite{Masip,NMSSM}. 
Moreover, $\lambda$ is required to be larger than $0.6$ to achieve 126 GeV Higgs mass with the s-top mass ($\widetilde{m}_t$) much lighter than 1 TeV, which is a necessary condition for the naturalness of the small Higgs mass. The relatively small value of $\lambda$ pushes ${\rm tan}\beta$ toward unity so that ${\rm sin}^22\beta$ [$=4{\rm tan}^2\beta/(1+{\rm tan}^2\beta)^2$] in the tree level correction of Eq.~(\ref{higgsmass}) becomes almost {\it maximized}. Thus, the perturbativity of $\lambda$ and the naturalness of the Higgs mass 
restrict the parameter space quite seriously \cite{nmssm2,HMRU}: 
\dis{
0.6\lesssim \lambda \lesssim 0.7 ~,\qquad 1\lesssim {\rm tan}\beta\lesssim 3 ~,
}
unless the ``maximal mixing scenario,'' which would also require a fine-tuning, is assumed. 
However, if the upper bound of $\lambda$ could somehow be relaxed even slightly to $0.8$-$1.0$ with its perturbativity maintained up to the GUT scale, ${\rm tan}\beta$ can be remarkably relieved to 4-8 for $\widetilde{m}_t=500$-$700$ GeV, yielding the 126 GeV Higgs mass \cite{HMRU}.\footnote{A strongly coupled region of $\lambda$ in energy scales ($\mu_-<\mu<\mu_+$) potentially ruins the gauge coupling unification. 
However, it could be reversely utilized to ameliorate the precision of the unification, 
only if $\gamma_{h_u,h_d}{\rm log}(\mu_+/\mu_-)$ 
are not too large for the strongly coupled region \cite{HMRU}. 
For other studies on non-perturbative $\lambda$ 
(by additional strong gauge interactions), 
see e.g. Refs.~\cite{strong}.  
}
\footnote{
For phenomenological analyses on cases with an order one $\lambda$ (``$\lambda$-SUSY''), see e.g. Refs.~\cite{Lsusy}.  
}      

From the Yukawa term Eq.~(\ref{Shh}), the beta function of $\lambda^2$ reads as the summation of the anomalous dimensions of $S$, $h_u$, and $h_d$:
\dis{ \label{beta}
\beta_{\lambda^2}=\mu\frac{d\lambda^2}{d\mu}=2\lambda^2\left(\gamma_S+\gamma_{h_u}+\gamma_{h_d}\right) .
}
While the Yukawa couplings make positive contributions to the anomalous dimensions and the beta function, the gauge couplings do negative ones to them. 
Thus, one can expect that the LP constraint is relaxed  by enhancing the gauge sector in which $S$, $h_u$, and $h_d$ are involved.
For a simple analysis, we will confine our discussion only on the case of a perturbative gauge interaction.
In this paper, we will attempt to relax the LP constraint by introducing new Abelian gauge symmetries, under which the MSSM Higgs, $\{h_u,h_d\}$ (and also the singlet $S$) are charged. Then the new gauge interactions would resist the blowup of $\lambda$ at higher energies.\footnote{Once a new gauge symmetry, under which the MSSM Higgs doublets are charged, is introduced, the D-term potential associated to it might be utilized to raise the tree-level Higgs mass. For a sizable effect by it, it turns out that the soft mass squareds of new Higgs, which break the new gauge symmetry, $\{\widetilde{m}_\phi^2, \widetilde{m}_{\phi^c}^2\}$ should be much heavier than the new gauge boson's mass squared \cite{Dterm}. Too large mass splittings of them could introduce a fine-tuning associated with the electroweak symmetry breaking \cite{Lodone}, which is a non-trivial constraint in this scenario. 
}

In this paper, {\it we intend to maintain the gauge coupling unification at the GUT scale}, which is one of the great achievements in the MSSM. 
In fact, SU(5) and SO(10) GUTs can provide the frameworks to extend the SM gauge group $G_{\rm SM}$ to a simple group, keeping the gauge coupling unification.  
Due to the gauge coupling unification and also doublet/triplet splitting in the Higgs sector, however, SU(5) and SO(10) should be broken around the GUT scale. 
Accordingly, one need to consider product gauge groups, $G_{\rm SM}\times G$.  

If the new gauge group $G$ is a non-Abelian group, the chiral matter sector of the MSSM as well as the Higgs sector are required to be extended by introducing more {\it chiral} fields  
such that they could be accommodated in non-trivial multiplets of $G$. 
Of course, the extra chiral matter should somehow be made heavy at low energy. 
%
%
With such other extra matter fields, however, 
the gauge coupling unification might be hard to be maintained, since extra matter fields would not be guaranteed to compose SU(5) or SO(10) multiplets at all.\footnote{If we give up the gauge coupling unification, one could find examples of the non-Abelian gauge extension. In Ref.~\cite{SU5^2}, a model of SU(3)$_c\times$SU(2)$_1\times$SU(2)$_2\times$U(1)$_Y$ is proposed to relax the LP constraint on $\lambda$. It is possibly embedded in a product group,   SU(5)$_1\times$SU(5)$_2$. The SU(3)$_c$ results from spontaneous symmetry breaking by a Higgs of the bi-fundamental representation under the two SU(3)s embedded in the two different SU(5)s. SU(2)$_L$ can be obtained in a similar way. The third family of the chiral matter and the Higgs are assumed to be charged only under SU(2)$_1$, while the first two families only under SU(2)$_2$. On the other hand, the singlet $S$ remains neutral.} 
%
%

Even in the case of a new Abelian gauge symmetry, however, anomaly cancellation conditions often require also the presence of extra matter fields, which have not been yet observed in the laboratories. Thus, the extra matter fields should be vector-like under the SM gauge symmetries such that they can obtain heavy masses below the breaking scale of $G$. Moreover, they should compose SU(5) or SO(10) multiplets for the gauge coupling unification. As mentioned above, in this paper we are interested in gauge extensions with Abelian groups in the NMSSM.  

In fact, the extra matter is helpful for relaxing the LP constraint of $\lambda$, only if they are embedded in SU(5) or SO(10) multiplets and made heavy at low energies: they would result in quite larger SM gauge 
couplings at higher energies compared to those in the original MSSM, 
and so enhance their negative contributions to the RG equation of $\lambda$. With five pairs of extra $\{{\bf 5},\overline{\bf 5}\}$, indeed, the allowed low energy value of $\lambda$ can be lifted to $0.8$, avoiding the LP below the GUT scale \cite{Masip, NMSSM}.

Our paper is organized as follows:  
we will survey promising U(1) gauge symmetries in section \ref{sec:U(1)s}.  
In section \ref{sec:RGanlys}, we will re-analyze the LP constraint in the presence of new U(1) gauge symmetries 
under various conditions. 
In section \ref{sec:model}, we will propose the concrete models reflecting the conditions considered in section \ref{sec:RGanlys}. 
In section \ref{sec:FCNC}, we will discuss phenomenological constraints on 
the breaking scale of the extra U(1) introduced in section \ref{sec:model}. 
Section \ref{sec:conclusion} will be devoted to conclusion.

\section{U(1) gauge symmetries embedded in the GUTs}
\label{sec:U(1)s}

In the search for anomaly-free combinations of extra matter under extra U(1) gauge symmetries, 
it is worthwhile to consider the U(1)s outside the SM gauge group $G_{\rm SM}$, but embedded in the well-known GUTs such as SO(10) and $E_6$: it can provide a guide for constructing a consistent model with gauge anomaly-free U(1)s.
But we will not discuss the GUTs themselves in this paper: {\it we just pragmatically employ such U(1)s embedded in the GUTs, particularly for easily obtaining anomaly-free matter contents.}  

The minimal GUT containing the SM gauge group is SU(5). 
Together with SU(5), U(1)$_X$ is embedded in SO(10).
Under the symmetry breaking ${\rm SO(10)\rightarrow  SU(5)\times U(1)}_X$, the spinor and vector representations of SO(10) are split as 
\dis{ \label{SO10split}
{\bf 16}\rightarrow {\bf 10}_{1/\sqrt{40}} + \overline{\bf 5}_{-3/\sqrt{40}} + {\bf 1}_{5/\sqrt{40}} ~,~~~~~~~{\bf 10}\rightarrow {\bf 5}_{-2/\sqrt{40}} + \overline{\bf 5}_{2/\sqrt{40}} ~, 
}
where the subscripts indicate the U(1)$_X$ charges. 
The above tensor and (anti-)fundamental representations of SU(5) can accommodate the following MSSM matter:
\dis{ \label{SO10matt}
&{\bf 10}_{1/\sqrt{40}}=\{u^c, q, e^c\}~,~~~\overline{\bf 5}_{-3/\sqrt{40}}=\{d^c, l\}~,~~~{\bf 1}_{5/\sqrt{40}} = \nu^c ~;
\\
&\qquad\quad~~ {\bf 5}_{-2/\sqrt{40}}=\{D, h_u\}~,~~~~\overline{\bf 5}_{2/\sqrt{40}}=\{D^c, h_d\} ~,
}
where the notation for the MSSM matter is self-evident. Throughout this paper, we will use the small (capital) letters for the MSSM (extra) superfields.   
Here $\{D,D^c\}$, whose MSSM quantum numbers are opposite to or same as $d^c$, are absent in the MSSM field spectrum unlike the other matter in \eq{SO10matt}. 
They spoil the gauge coupling unification, and possibly lead to too fast proton decay. 
Thus, they should be dropped from the field content of the low energy effective theory.
Even without $\{D,D^c\}$,  
the anomaly-free conditions associated with U(1)$_X$ are still fulfilled, 
since they are exactly vector-like under ${\rm G_{SM}\times U(1)}_X$.
Not only $\{D,D^c\}$ but also the MSSM Higgs $\{h_u,h_d\}$ carry the opposite charges $\mp 2/\sqrt{40}$. On the other hand, the singlet $S$ in Eq.~(\ref{Shh}) still remains neutral. 
Since the Higgs in Eq.~(\ref{Shh}) are charged under U(1)$_X$, the U(1)$_X$ gauge symmetry 
could be helpful for relaxing the LP constraint on $\lambda$.   

Like the U(1)$_X$, the U(1)$_{B-L}$ symmetry also resides between SO(10) and $G_{\rm SM}$, even if it is not ``orthogonal'' to $G_{\rm SM}$. 
Under U(1)$_{B-L}$, however, the Higgs $\{h_u,h_d\}$ as well as the singlet $S$ are neutral, and so it is not useful for relaxing the LP constraint. The U(1) symmetry embedded in the SU(2)$_R$ of the Pati-Salam gauge group, SU(4)$_c\times$SU(2)$_L\times$SU(2)$_R$ is also an interesting gauge group. However, it can be obtained just by a linear combination of U(1)$_X$ and U(1)$_{B-L}$.  

Another interesting U(1) symmetry is ``U(1)$_Z$''  embedded in $E_6$ together with SO(10).   
Under the symmetry breaking, $E_6\rightarrow  {\rm SO(10)\times U(1)}_Z$, the fundamental representation of $E_6$, ${\bf 27}$ is split as follows: 
\begin{eqnarray} \label{E6split}
{\bf 27}\rightarrow {\bf 16}_{1/\sqrt{24}} + {\bf 10}_{-2/\sqrt{24}} +{\bf 1}_{4/\sqrt{24}} ~,
\end{eqnarray}
where the subscripts denote the charges of U(1)$_Z$. 
Hence, one ${\bf 27}$ contains one family of the SM chiral matter, one Higgs doublet pair, extra colored matter $\{D,D^c\}$, and a singlet: 
\dis{ \label{E6matt}
{\bf 16}_{1/\sqrt{24}} =\{u^c, q, e^c~;~ d^c, l~;~ \nu^c\}~,~~~
{\bf 10}_{-2/\sqrt{24}}=\{D, h_u ~;~ D^c, h_d\} ~, ~~~
{\bf 1}_{4/\sqrt{24}}= S ~, 
}
As a result, if one introduces three families of ${\bf 27}$, two more pairs of the Higgs doublets and three $\{D, D^c\}$, as well as three singlets in total should be accompanied with the MSSM chiral matter. 
Note that $S$ in Eq.~(\ref{E6matt}) can be the NMSSM singlet appearing in Eq.~(\ref{Shh}), since it has the U(1)$_Z$ charge of $4/\sqrt{24}$, while $h_u$ and $h_d$ both carry $-2/\sqrt{24}$. Accordingly, U(1)$_Z$ can also be helpful for avoiding the LP for $\lambda$ below the GUT scale. According to our analysis, U(1)$_Z$ turns out to be much more efficient than U(1)$_X$ in relaxing the LP constraint. Thus, we will mainly focus on U(1)$_Z$. 

For a gauge interaction of $G$ to efficiently seize $\lambda$ in the perturbative regime, the following conditions should be generically satisfied: 

{\bf (1)} The gauge coupling associated with $G$ needs to be large enough at the GUT scale. 

{\bf (2)} The beta function coefficient of $G$ needs to be small enough.

{\bf (3)} The breaking energy scale of $G$ should be low enough.   

\noindent
%
In order to reflect the condition {\bf (1)} in the model, one would not require that the gauge couplings of U(1)$_Z$ and U(1)$_X$ are necessarily unified with the SM gauge couplings at the GUT scale: relatively larger U(1)$_{Z,X}$ gauge couplings than those of the MSSM is allowed.
Alternatively, we can take the U(1)$_{Z}$  [U(1)$_X$] charge normalization smaller than the $E_6$ [SO(10)] normalization ``$\sqrt{24}$'' [``$\sqrt{40}$''], assuming that the U(1)$_Z$ [U(1)$_X$] gauge coupling is unified with the SM gauge couplings at the GUT scale. 
We just naively anticipate that such a charge normalization of U(1)$_Z$ and U(1)$_X$ can be supported by a proper UV theory embedding our model. 
Of course, both yield the same result in lifting $\lambda$. 
Throughout this paper, we take the latter choice.   

For fulfilling the condition {\bf (2)}, we will assign the U(1)$_Z$ and U(1)$_X$ charges to only one family of the MSSM chiral matter and one pair of the Higgs doublets together with $\{D,D^c\}$ and $S$, which compose an anomaly-free combination of the matter. Hence, the other two families of the chiral matter cannot couple to the Higgs, because they remain neutral under U(1)$_Z$ and U(1)$_X$, while the Higgs doublets carry the charges.  
Thus, we should introduce additional vector-like matter such that the desired Yukawa couplings for them can be generated after U(1)$_Z$ and U(1)$_X$ breakings. 
Since the extra matter $\{D,D^c\}$ can mediate unwanted too fast proton decay, we need to introduce 
a (global) symmetry in order to forbid such a possibility. 
 
If the U(1)$_Z$ breaking scale is too low,   
the condition {\bf (3)} can be in conflict with the  constraints on FCNC processes, since our U(1)$_Z$ and U(1)$_X$ charge assignments are family-dependent. Moreover, too low breaking scale of U(1)$_Z$ could affect also the precision tests of the SM gauge interactions associated with the ``$S$'' and ``$T$'' parameters. Hence, we take the U(1)$_Z$ and U(1)$_X$ breaking scales of 5-10 TeV.  

In principle, models with such modified U(1) gauge symmetries could originate from a GUT defined in a higher dimensional spacetime such as the heterotic string theory compactified in an orbifold \cite{Horbifold}, in which U(1)$_{Z,X}$ gauge symmetries are embedded in $E_8\times E_8^\prime$. 
However, the discussion on it would be beyond the scope of this paper.  

\section{Lifting the $\lambda$ coupling constant} \label{sec:RGanlys}

In this paper, we will consider only the SM gauge group $G_{\rm SM}$ and the U(1)$_Z$ and U(1)$_X$, which are motivated by the U(1)s embedded in $E_6$ and SO(10). For a small beta function coefficients $b_{Z,X}$, we assign the U(1)$_{Z,X}$ charges only to one family of the MSSM chiral matter (the third family of the quarks, leptons, and their superpartners) and the Higgs, introducing the extra colored matter $\{D,D^c\}$ and a singlet $S$. 
In Table \ref{tab:Qnumb1}, we present their charges  of the {\it global} U(1)$_R$ as well as the gauged U(1)$_Z$ and U(1)$_X$.  
\begin{table}[!h]
\begin{center}
\begin{tabular}
{c|ccc|cccccc|cccc} 
{\rm Superfields}  &   $S$   &
 $h_u$  &  $h_d$  &  $u^c_3$  &
 $q_3$  &  $e^c_3$  &  $d^c_3$  &  $l_3$  &  $\nu^c_3$ 
 &  $D$  &  $D^c$  &  $H_u$  &  $H_d$ 
  \\
\hline 
U(1)$_Z$ ($\times n_Z/\sqrt{24}$)  & ~$4$ & $-2$ & $-2$ & ~$1$
 & ~$1$ & ~$1$ & ~$1$ & ~$1$ & ~$1$ & $-2$ & $-2$ & ~$0$ & ~$0$ \\
 U(1)$_X$ ($\times n_X/\sqrt{40}$)  & ~$0$ & $-2$ & ~$2$ & ~$1$
 & ~$1$ & ~$1$ & $-3$ & $-3$ & ~$5$ & $-2$ & ~$2$ & ~$0$ & ~$0$ \\
U(1)$_{R}$ & $-2$ & ~$2$ & ~$2$ & $-1$ & ~$1$ & ~$1$ & $-1$ & $-1$ & ~$2$ & ~$1$ & $-1$ & ~$0$ & ~$0$
\\
\end{tabular}
\end{center}\caption{Charge assignments of the gauged U(1)$_Z$, U(1)$_X$, and the global U(1)$_R$. A singlet, the MSSM Higgs doublets, the third family of the chiral matter, and an extra vector-like pair $\{D,D^c\}$ carry the charges. They compose anomaly-free matter contents under $G_{\rm SM}\times$U(1)$_Z\times$U(1)$_X$. 
$\{H_u,H_d\}$ are necessary for the Yuakawa couplings for the first and second families of the chiral matter, and also gauge coupling unification. 
}\label{tab:Qnumb1}
\end{table}
The charges for the third family of the MSSM chiral matter in Table \ref{tab:Qnumb1} just follow the charge assignments of Eqs.~(\ref{E6matt}) and (\ref{SO10matt}). 
Concerning the $E_6$ and SO(10) charge normalizations, 
refer to e.g. Ref.~\cite{flipF}. 
Since they compose a {\bf 27} of $E_6$, all the gauge anomalies must be canceled out. Note that $\{h_u, h_d\}$ and $\{D, D^c\}$ are vector-like under $G_{\rm SM}\times $U(1)$_X$, but not under U(1)$_Z$.
Even if the U(1)$_Z$ and U(1)$_X$ embedded in the GUTs are introduced, we don't follow the charge normalization determined when they are embedded in the GUTs. Thus, $n_Z$ and $n_X$ are not rigorously required to be unity in our case. 
For the SM gauge coupling unification, 
two lepton doublets $\{H_u,H_d\}$ should be supplemented as seen in Table \ref{tab:Qnumb1}, 
which do not have any charges of U(1)$_Z\times$U(1)$_X$. 
          
The charge assignments of U(1)$_Z\times$U(1)$_X$ and U(1)$_R$ in Table \ref{tab:Qnumb1} permit the Yukawa couplings for the third family of the chiral matter and the Higgs doublets: 
\dis{ \label{W3}
W_3=\lambda Sh_uh_d + y_tq_3h_uu^c_3 + y_bq_3h_dd^c_3 
+ y_\tau l_3h_de^c_3 ~,
}
in which the MSSM $\mu$ term is promoted to the trilinear $\lambda$ coupling. 
Throughout this paper, we suppose that $y_b$ and $y_\tau$ are relatively smaller than $\lambda$ and $y_t$. 
Note that $\nu_3^c$ cannot couple to the lepton doublet $l_3$ and the Higgs $h_u$ due to the U(1)$_R$ symmetry. We assume that it develops a VEV of order TeV scale, breaking U(1)$_Z$ completely. 
%
%
%

The mass squared of $h_u$ is assumed to be negative at low energies, and so it 
leads to a non-zero VEV of the Higgs, 
triggering the electroweak symmetry breaking. Note that we have an additional quartic term $\lambda^2|h_uh_d|^2$ in the scalar potential apart from the quartic potential coming from the ``D-term.'' 
So there is no D-flat direction in the D-term potential of the Higgs unlike the MSSM.  
The ``A-term'' corresponding to the $\lambda$ term in Eq.~(\ref{W3}) provides a tadpole of $\widetilde S$, and so a VEV of $\widetilde S$ 
can also be developed. 
It could induce the MSSM ``$\mu$'' parameter ($\mu_{\rm eff}\equiv \lambda\langle \widetilde S\rangle$), which is a SUSY mass parameter of $\{h_u,h_d\}$.

If $\langle \widetilde S\rangle$ is the main source of the U(1)$_Z$ breaking, it should be large enough to avoid low energy constraints on an extra U(1). Then, $\lambda$ should be small enough to ensure the small Higgs and Higgsino masses. 
However, a small $\lambda$ cannot enhance the quartic coupling of the Higgs potential, and so a sizable fine-tuning becomes unavoidable \cite{King}. 
Hence, we will assume $\langle \widetilde S\rangle \sim {\cal O}(1)$ TeV or lower, and introduce another main breaking source of U(1)$_{Z}$ separately.    

Since there is no $S^3$ term in the superpotential \eq{W3} unlike the ordinary NMSSM, one might think that there exists an accidental Pecci-Quinn (PQ) symmetry. However, such a global symmetry, under which $S$ carries a non-zero charge, is gauged to U(1)$_Z$ in this case. Even below the U(1)$_Z$ breaking scale, which is assumed to be higher than $\langle S\rangle$, we will show later the absence of such an accidental PQ symmetry. 

At the moment, $\{D,D^c\}$ and $\{H_u,H_d\}$ remain massless because of the U(1)$_R$.  
We will explain how they get their masses in section \ref{sec:model}. 
As will be seen later, they and other vector-like fields play important roles to induce the ordinary Yukawa couplings for the first and second families of the chiral matter.


Although we have not yet proposed a full model with U(1)$_Z\times$U(1)$_X$ charge assignments, we first attempt to perform a relatively model independent analysis on how much the $\lambda$ coupling in Eq.~(\ref{W3}) can be lifted at low energy. 
We will discuss the cases of U(1)$_Z$ [Case I] and U(1)$_Z\times$U(1)$_X$ [Case II]. 

The solution of the RG equation for the three MSSM gauge couplings are given by 
\begin{eqnarray}
g_k^2(t)=\frac{g_U^2}
{1+\frac{g_U^2}{8\pi^2}b_k(t_0-t)} \qquad {\rm for} ~~k=3,~2,~1,
\end{eqnarray}
where $t$ parametrizes the renormalized mass scale, 
$t={\rm log}(\mu/M_{\rm GUT})$. $b_k$ ($k=3,2,1$) denotes the beta function coefficients of the gauge couplings for SU(3)$_c$, SU(2)$_L$ and U(1)$_Y$. 
In the presence of the extra $v$ pairs of $\{{\bf 5},\overline{\bf 5}\}$, they are given by $b_k=(-3+v,1+v,33/5+v)$, where $v=0$ corresponds to the case of the MSSM.
For the matter content of Table \ref{tab:Qnumb1} ($v=1$), the unified gauge coupling $g_U^2$ is estimated as $0.62$. 
With the one more extra pair of $\{{\bf 5},\overline{\bf 5}\}$ ($v=2$), $g_U^2$ is lifted to $0.82$. 
For $v=3$, $4$, and $5$, $g_U^2$ are given by $1.18$, $2.13$, and $11.19$, respectively.

Similarly, the solution to the RG equations of the U(1)$_Z$ and U(1)$_X$ gauge couplings are 
\begin{eqnarray}  \label{gZgX}
g_Z^2(t)=\frac{g_{Z0}^2}
{1+\frac{g_{Z0}^2}{8\pi^2}b_Z(t_0-t)} ~~ {\rm for}~~ t>t_Z ~;\quad 
g_X^2(t)=\frac{g_{X0}^2}
{1+\frac{g_{X0}^2}{8\pi^2}b_X(t_0-t)} ~~ {\rm for}~~ t>t_X , 
\end{eqnarray}
where $b_{Z,X}$ indicates the beta function coefficient of U(1)$_{Z,X}$, 
and $t_{Z,X}$ parametrizes the U(1)$_{Z,X}$ breaking scale $M_{Z,X}$ [$t_{Z,X}\equiv {\rm log}(M_{Z,X}/M_{\rm GUT})$].
In particular, Case I corresponds to the case of setting $g_{X0}^2=0$, which turns off the U(1)$_X$ gauge interaction.  
Only with the field contents in Table \ref{tab:Qnumb1}, the beta function coefficients $b_{Z,X}$ are given by $3n_{Z,X}^2$. 
As will be discussed later, however, some additional vector-like superfields charged under U(1)$_Z$ [and U(1)$_X$] are necessary in order to induce the desired Yukawa couplings for the first two families of the matter in Case I [Case II]. Then, $b_Z$ becomes $(4+\frac{1}{12})n_Z^2$ in Case I  [$(4+\frac12)n_Z^2$ in Case II].
With such additional matter in Case II, $b_X$ would increase to $(4+\frac{3}{10})n_X^2$.  
Note that    
$b_{Z,X}$ in both cases are quite smaller than those when all the three families of the matter and additional Higgs are charged, $3\times 3n_{Z,X}^2$. 
Since we don't consider the GUT of $E_6$ or SO(10), 
the normalization factors $n_{Z,X}$ in Table \ref{tab:Qnumb1} and in $b_{Z,X}$ of \eq{gZgX} don't have to be unity in our case. Instead, we set $g_{Z0}^2=g_{X0}^2=g_{U}^2$. In the perturbative regime, $(n_{Z,X}g_{Z,X})^2$, which are effective gauge couplings, should be quite smaller than $4\pi^2$. 

With the charge assignments in Table \ref{tab:Qnumb1} and the Yukawa interactions in \eq{W3}, one can obtain the 1-loop anomalous dimensions for $S$, $h_u$, $h_d$, $q_3$ and $u^c_3$ in the standard manner: 
\begin{eqnarray}
&&16\pi^2\gamma_S= 2\lambda^2 -\frac{4}{3}(n_Zg_Z)^2 , \\
&&16\pi^2\gamma_{h_u}= \lambda^2+3y_t^2-\frac{3}{2}g_2^2-\frac{3}{10}g_1^2-\frac{1}{3}(n_Zg_Z)^2 -\frac{1}{5}(n_Xg_X)^2 , \\
&&16\pi^2\gamma_{h_d}= \lambda^2 -\frac{3}{2}g_2^2-\frac{3}{10}g_1^2-\frac{1}{3}(n_Zg_Z)^2 -\frac{1}{5}(n_Xg_X)^2 , \\
&&16\pi^2\gamma_{q_3}= y_t^2-\frac{8}{3}g_3^2-\frac{3}{2}g_2^2-\frac{1}{30}g_1^2-\frac{1}{12}(n_Zg_Z)^2 -\frac{1}{20}(n_Xg_X)^2 , \\
&&16\pi^2\gamma_{u^c_3}= 2y_t^2-\frac{8}{3}g_3^2-\frac{8}{15}g_1^2-\frac{1}{12}(n_Zg_Z)^2 -\frac{1}{20}(n_Xg_X)^2 , 
\end{eqnarray}
where we ignored the contributions by the Yukawa couplings of $y_b$ and $y_\tau$.
The $(n_Zg_Z)^2$ and $(n_Xg_X)^2$ terms in the above  anomalous dimensions, which are all negative, result from the U(1)$_Z$ and U(1)$_X$ gauge interactions, respectively. 
Then, it is straightforward to write down the RG equation for the $\lambda$ and $y_t$ couplings: 
\begin{eqnarray} \label{RGeq}
&&\quad~~ 
\frac{d\lambda^2}{dt}
=\frac{\lambda^2}{8\pi^2}\left[4\lambda^2+3y_t^2-3g_2^2
-\frac35g_1^2-2(n_Zg_Z)^2 -\frac{2}{5}(n_Xg_X)^2\right] , \\ \label{RGeq2}
&& \frac{dy_t^2}{dt}
=\frac{y_t^2}{8\pi^2}\left[\lambda^2+6y_t^2-\frac{16}{3}g_3^2-3g_2^2-\frac{13}{15}g_1^2-\frac{1}{2}(n_Zg_Z)^2 -\frac{3}{10}(n_Xg_X)^2\right] . 
\end{eqnarray}
Based on the analyses with Eqs.~(\ref{RGeq}) and (\ref{RGeq2}), we display the upper bounds of $\lambda$ ($\equiv\lambda_{\rm max}$) at low energy for Case I in Fig. 1-(a), and Case II in Fig 1-(b), respectively. In these analyses, we naively assume that the gauged U(1)s' breaking scales are around 5 TeV, and the masses of the extra multiplets are 5-10 TeV. Below 5 TeV, thus, we ignored the $g_{Z,X}$'s contributions to Eqs.~(\ref{RGeq}) and (\ref{RGeq2}).

\begin{figure}
\begin{center}
\subfigure[]
{\includegraphics[width=0.48\linewidth]{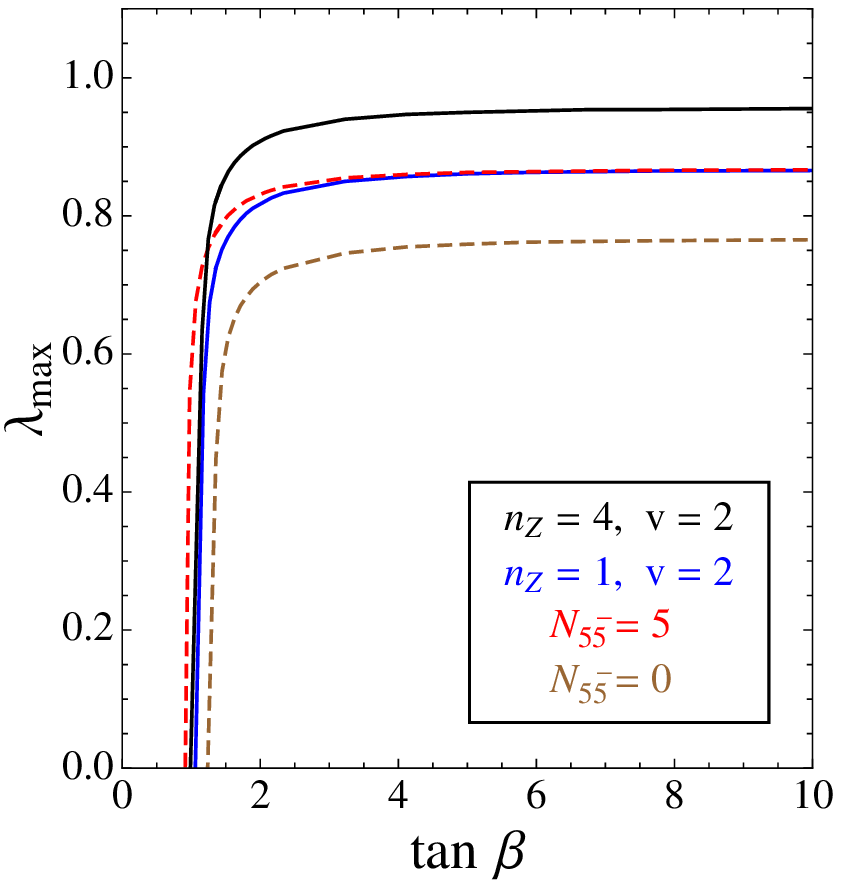}}
\hspace{0.2cm}
\subfigure[] 
{\includegraphics[width=0.48\linewidth]{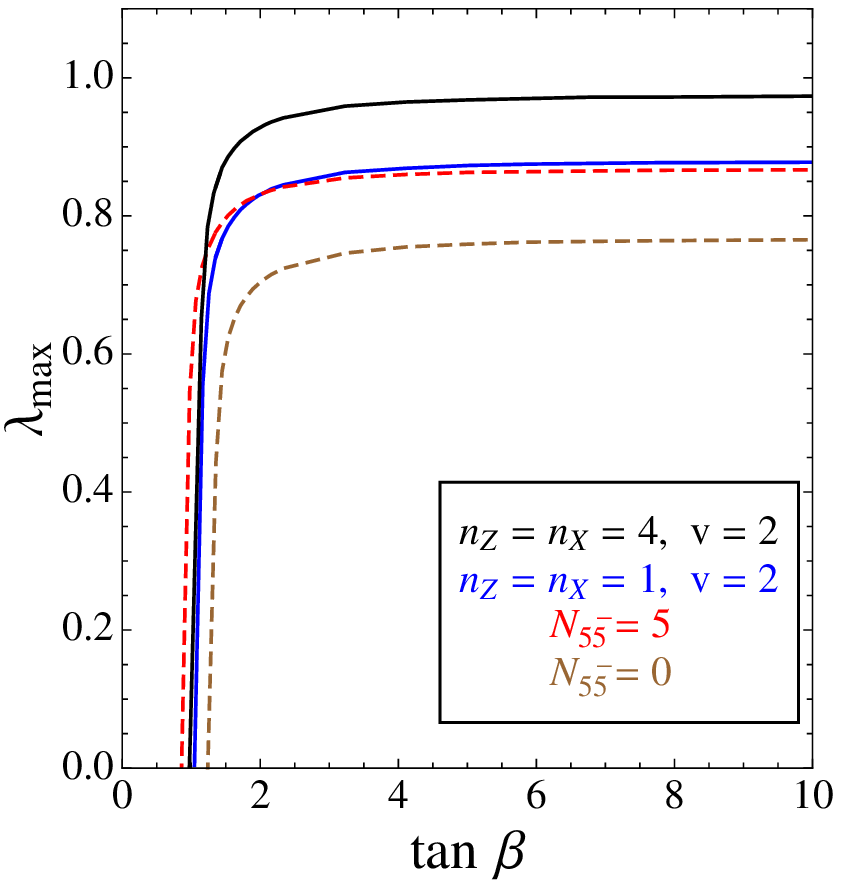}}
\end{center}
\caption{$\lambda_{\rm max}$ vs. tan$\beta$ for {(a)} the U(1)$_Z$ extension (Case I), and {(b)} the  U(1)$_Z\times$U(1)$_X$ extension (Case II) of the NMSSM.  
The $n_{Z}=1$ [$n_X=1$] corresponds to the case that the charge normalization of U(1)$_Z$ [U(1)$_X$] determined by $E_6$ [SO(10)] is employed.  
$v$ stands for the number of the extra $\{{\bf 5},\overline{\bf 5}\}$, which are charged under U(1)$_Z$ [and also U(1)$_X$], while 
$N_{5\bar{5}}$ the total numbers of extra $\{{\bf 5},\overline{\bf 5}\}$ in the absence of the extra U(1)s. 
In both cases, the U(1) breaking scales are set to be 5 TeV. 
In Case II, $\lambda _{\rm max}$ slightly more increases compared to Case I.
}
\end{figure}

As seen in Fig. 1-(a), $\lambda_{\rm max}$ in the absence of extra U(1)s and matter is given by about $0.7$. If five pairs of $\{{\bf 5},\overline{\bf 5}\}$ are added, $\lambda_{\rm max}$ increase up to $0.8$ in the absence of the extra U(1)s \cite{Masip}. We note the similar result, $\lambda_{\rm max}\approx 0.8$ can be achieved also in Case I [i.e. when only U(1)$_Z$ is turned on] with the two pairs of $\{{\bf 5},\overline{\bf 5}\}$, even if the $E_6$ normalization ($n_Z=1$) is employed. For $n_Z=4$, which is almost the maximal possibility in the perturbative regime [$(n_Zg_U)^2\approx (3.62)^2$], $\lambda_{\rm max}$ can reach $0.95$ for ${\rm tan}\beta >3$. 
When both U(1)$_Z$ and U(1)$_X$ are turned on, $\lambda_{\rm max}$ can be slightly more lifted: Fig. 1-(b) shows that it becomes close to $1.0$ ($0.9$) for ${\rm tan}\beta >3$ with $n_{Z}=n_{X}=4$ ($n_{Z}=n_{X}=1$). 

In fact, two-loop corrections start being sizable in the existence of three or more pairs of $\{{\bf 5},\overline{\bf 5}\}$ \cite{U1nmssm}. 
Moreover, the case with five pairs of $\{{\bf 5},\overline{\bf 5}\}$ is the perturbatively marginal case, since the expansion parameter associated with the MSSM gauge couplings, $g_U^2/4\pi$ becomes $0.89$ at the GUT scale. 
Even in the case of two pairs of $\{{\bf 5},\overline{\bf 5}\}$ we considered in Fig. 1, 
$g_Z$ and $\lambda$ reach the perturbatively marginal values at the GUT scale. 
In order to get the upper bound on $\lambda$ in the absence or presence of the extra U(1)s, however, 
we considered such extreme cases in Fig. 1.
For more precise results, more rigorous estimations including two-loop corrections would be needed. 

\section{The Models}
\label{sec:model}

While the Yukawa couplings for the third family of the MSSM chiral matter are allowed as seen in Eq.~(\ref{W3}), the first and second families of the MSSM matter fields cannot yet 
couple to the ordinary MSSM Higgs doublets $\{h_u, h_d\}$, since they don't carry U(1)$_Z$ and U(1)$_X$ charges unlike the Higgs doublets. 
%
%
Moreover, $\{D,D^c\}$ and $\{H_u,H_d\}$ should have  mass terms. 
Thus, we need to introduce some vector-like superfields in order to induce all the desired Yukawa couplings and mass terms. Now we propose the two models as examples.

\underline{\bf Model of U(1)$_Z$}
In the first model, we consider only the gauge U(1)$_Z$ and the global U(1)$_R$ symmetries. Thus, we ignore the U(1)$_X$ charges of Table \ref{tab:Qnumb1} in the first model. 
The charge assignments for the first and second families of the MSSM chiral superfields and other needed vector-like matter are listed in Table \ref{tab:Qnumb2}.
\begin{table}[!h]
\begin{center}
\begin{tabular}
{c|ccccc|cccc|c} 
{\rm Superfields}  &   $X$   &
 $Y$  &  $Y^c$  &  $Z$  &  $Z^c$   &  $D^{\prime}$   &  $D^{c\prime}$  &  $L$  &  $L^c$  &  $u^c_{1,2}$ ~  
 $q_{1,2}$ ~  $e^c_{1,2}$ ~  $d^c_{1,2}$ ~  $l^c_{1,2}$ ~  $\nu^c_{1,2}$ 
\\
\hline 
U(1)$_Z$ & ~$0$  &  ~$1$  &  $-1$ & ~$2$ & $-2$  & ~$0$ & ~$0$ & ~$2$ & $-2$ & ~$0$
  \\
U(1)$_{R}$ & ~$0$ & ~$0$ & ~ $0$ & ~$0$ & ~$0$ & ~$3$ & ~$1$ & $-1$ & ~$1$ & ~$1$  
\\
\end{tabular}
\end{center}\caption{Charge assignments for the gauge U(1)$_Z$ and global U(1)$_R$ for the extra vector-like fields and the first and second families of the chiral matter. Here the U(1)$_Z$ charge normalization, $n_Z/\sqrt{24}$ is omitted for a simple presentation.  
The first and second families of the ordinary chiral matter do not carry U(1)$_Z$ charges. Since $\{D^\prime,D^{c\prime}\}$ and $\{L,L^c\}$ compose $\{{\bf 5},\overline{\bf 5}\}$, they maintain the gauge coupling unification. $X$, $\{Y,Y^c\}$, and $\{Z,Z^c\}$ are inert under the SM gauge interactions.  
}\label{tab:Qnumb2}
\end{table}
$X$, $\{Y,Y^c\}$, and $\{Z,Z^c\}$ in Table \ref{tab:Qnumb2} are the MSSM singlet superfields. $\{D^\prime,D^{c\prime}\}$ and $\{L,L^c\}$ are extra SU(3)$_c$ triplets and SU(2)$_L$ doublets.   
Since $\{D^\prime, D^{c\prime}\}$ and $\{L,L^c\}$ are embedded in a pair of $\{{\bf 5},\overline{\bf 5}\}$, 
the gauge coupling unification can still be maintained. 
With the field contents in Table \ref{tab:Qnumb1} and \ref{tab:Qnumb2}, the beta function coefficient of U(1)$_Z$ is estimated as $b_Z=(4+\frac{1}{12})n_Z^2$. 
It was utilized for analyses of the $v=2$ cases in Fig. 1-(a). 

The relevant superpotential for the first two families of the MSSM chiral matter and $\{Z,Z^c\}$ are written as follows: 
\dis{ \label{W12}
W_{1,2}&=\sum_{i,j=1,2}\left(y^{ij}_uq_iu^c_jH_u + y^{ij}_dq_id^c_jH_d + y^{ij}_{\nu}l_i\nu^c_jH_u + y^{ij}_{e}l_ie^c_jH_d + M^{ij}_{\nu}\nu^c_i\nu^c_j \right)
\\
&\quad + Z\left(y_{h1}H_uh_d +  y_{h2}H_dh_u + y_{D1}DD^{c\prime} + y_{D2}D^cD^{\prime} 
\right) +y_V\nu_3^cYZ^c ,  
}
where we assume the Yukawa coupling constants, $y_{h1}$ and $y_{h2}$ are small enough to guarantee the light enough Higgs mass.  
We will discuss later how Eq.~(\ref{W12}) should be modified when U(1)$_X$ is also introduced.
The soft SUSY breaking ``A-term'' corresponding to the $y_V$ term in the scalar potential ($\equiv y_VA_V\tilde{\nu}_3^c\widetilde{Y}\widetilde{Z}^c)$ provides tadpoles of $\nu_3^c$, $Y$, and $Z^c$, and so it  
can generate non-zero VEVs of them. 

Indeed, the VEVs of $\{\tilde{\nu}_3^c, \widetilde{Y}, \widetilde{Z}^c\}$ can be of order $m_{\rm soft}/y_V$. 
With a relatively smaller $y_V$, thus, we get higher energy scale VEVs for them than the typical soft mass scale. 
It is possible, basically because the quartic terms in the scalar potential, which makes the scalar potential  bounded from below, come from the $y_V$ term in \eq{W12} with the coefficient of $|y_V|^2$. 
With the $y_V$ term in the superpotential \eq{W12} and the D-term potential ($\equiv\frac{g_Z^2}{2}D_Z^2$), one can show that $\{\tilde{\nu}_3^c, \widetilde{Y}, \widetilde{Y}^c, \widetilde{Z}, \widetilde{Z}^c\}$ should satisfy the following conditions at the minimum: 
\dis{
&~~\widetilde{Y}^c =0 ~, 
~~2g_Z^2D_{Z}=2g_Z^2\left(|\tilde{\nu}_3^c|^2+|\widetilde{Y}|^2-|\widetilde{Y}^c|^2+2|\widetilde{Z}|^2-2|\widetilde{Z}^c|^2\right)=-\widetilde{m}_Z^2 ~,
\\ 
&y_V^2\left(|\tilde{\nu}_3^{c}|^2+|\widetilde{Y}|^2
+|\widetilde{Z}^{c}|^2\right)
=\frac{4{\rm cot}^2\theta(\widetilde{m}_Z^2+\widetilde{m}_{Z^c}^2)
-{\rm sin}^22\phi(\widetilde{m}_{\nu_3^c}^2+\widetilde{m}_Y^2-\widetilde{m}_Z^2)}
{(1+{\rm cos}^2\theta) ~{\rm sin}^2 2\phi-4{\rm cos}^2\theta} ~,
}
Here $\widetilde{m}_{\nu_3^c}^2$, $\widetilde{m}_Y^2$, $\widetilde{m}_Z^2$ and $\widetilde{m}_{Z^c}^2$ 
indicate the soft mass squareds of $\tilde{\nu}_3^c$,  $\widetilde{Y}$, $\widetilde{Z}$, and $\widetilde{Z}^c$, respectively, and $\theta$ and $\phi$ parametrize $\tilde{\nu}_3^c$, $\widetilde{Y}$, and $\widetilde{Z}^c$ as follows:
\dis{
|\tilde{\nu}_3^{c}|\equiv R~{\rm sin}\theta~{\rm cos}\phi ~, 
~~|\widetilde{Y}|\equiv R~{\rm sin}\theta~{\rm sin}\phi ~, ~~|\widetilde{Z}^c|\equiv R~{\rm cos}\theta
}
where $R\equiv\sqrt{|\widetilde{\nu}_3^{c}|^2+|\widetilde{Y}|^2
+|\widetilde{Z}^{c}|^2}$. With smaller $y_V$, hence, the VEVs of $\{\tilde{\nu}_3^c, \widetilde{Y}, \widetilde{Z}^c\}$ can be larger than the typical sizes of the soft mass parameters. Note that the sign of $\widetilde{m}_Z^2$ can be negative at low energies, if $y_{D{1,2}}$ in \eq{W12} is of order unity. [Since $\{D,D^{c\prime}; D^c,D^\prime\}$ are colored particles, $y_{D{1,2}}$ of order one 
would not incur another LP problem.]    
We suppose that the VEVs for $\{\tilde{\nu}_3^c, \widetilde{Y}, \widetilde{Z}^c\}$  are around 5-10 TeV, breaking U(1)$_X$ at that scale, even if the typical soft masses are assumed to be of order TeV. 
%
For instance, $\tilde{\nu}_3^c\approx 6.4$ ($11.8$) TeV, $\tilde{Y}\approx 5.7$ ($9.8$) TeV,  $\widetilde{Z}\approx 7.1$ ($6.9$) TeV, $\widetilde{Z}^c\approx 9.3$ ($12.9$) TeV, for $y_V=0.1$, $g_Z^2=0.3$, $\phi=41.6^{\rm o}$ ($39.7^{\rm o}$), $\theta=42.5^{\rm o}$ ($50.1^{\rm o}$), $A_V=-2.6$ ($-3.4$) TeV, and $\sqrt{\widetilde{m}_{\nu_3^c}^2
-\frac12\widetilde{m}_Z^2}=1$ TeV, $\sqrt{\widetilde{m}_{Y}^2-\frac12\widetilde{m}_Z^2}=1.2$ ($1.5$) TeV, $\sqrt{\widetilde{m}_{Z^c}^2-\frac12\widetilde{m}_Z^2}=0.6$ ($0.8$) TeV, respectively.


$\{H_u, H_d\}$ are assumed to be relatively (say, about five times) heavier than other superpartners of the SM chiral fermions.
It is possible because a ``$\mu$ term'' for them, $\mu_HH_uH_d$ can be induced in the superpotential, e.g. via the Giudice-Masiero mechanism \cite{GM}, as will be shown later. 
$\{H_u, H_d\}$ can be integrated out to yield the effective SM Yukawa couplings, since they are much heavier than the SM chiral fermions and Higgs. 
%
After integrating out $\{H_u,H_d\}$, 
thus, the Yukawa couplings for the masses of the first and second families of the SM chiral fermions can be generated, which are estimated as 
$y^{ij}_{d,e;u,\nu}y_{h1;2}\langle \widetilde Z\rangle/\mu_H$. 
Note that 
they still perturbatively consistent for small enough dimensionless couplings $y^{ij}_{d,e;u,\nu}$ ($i,j=1,2$) and $y_{h1;2}$, even if $\langle \widetilde Z\rangle/\mu_H\sim {\cal O}(1)$. 

Once $\widetilde{Z}$ and $\{h_u,h_d\}$ develop VEVs, 
$\{H_u,H_d\}$ also acquire VEVs through the ``A-term'' corresponding to the $y_{h1}$ and $y_{h2}$ terms in \eq{W12}. 
$\langle H_{u,d}\rangle$ are estimated as $y_{h1,2}m_{3/2}\langle\widetilde{Z}\rangle \langle h_{d,u}\rangle/\mu_H^2$, which are much suppressed than $\langle h_{u,d}\rangle$. 
One should note here that the VEVs, $\langle H_{u,d}\rangle$ are along the directions of $\langle h_{d,u}\rangle$. 
Accordingly, $\langle H_{u,d}\rangle$ keep intact the electromagnetic U(1) gauge symmetry.  
All the charged components in $\{H_{u},H_d\}$ get  heavy masses from the soft mass terms, the $\mu_HH_uH_d$ term and its corresponding ``B$\mu$-term.''          

As mentioned in section \ref{sec:RGanlys}, we assume that $\langle \widetilde S\rangle$ is small enough [${\cal O}(1)$  TeV or lower ($\ll \langle \widetilde Z\rangle$)] that the effective $\mu$ ($\equiv\lambda\langle \widetilde S\rangle$) is not much large. 
Hence, one might expect an accidental PQ symmetry below the scale of $\langle \widetilde Z\rangle$, which is regarded as the dominant U(1)$_Z$ breaking source.
After $\{H_u,H_d\}$ decoupled, however, a bare $\mu$ term is also induced: 
\dis{
\left(y_{h1}y_{h2}\frac{\langle \widetilde Z\rangle^2}{\mu_H}\right)h_uh_d , 
}    
and so there does not remain an accidental PQ symmetry below the U(1)$_Z$ breaking scale. 

Note that the new vector-like colored particles $\{D^{\prime}, D^{c\prime}\}$ introduced in Table \ref{tab:Qnumb2} and Eq.~(\ref{W12}) couple to $\{D,D^c\}$ of Table \ref{tab:Qnumb1} and get masses, when $\widetilde Z$ develops a VEV of order 10 TeV scale. 
%
In fact, the quantum numbers of $d^c_{1,2}$ are the same as those of $D^{c\prime}$ [neglecting U(1)$_X$], and so $d^c_{1,2}$ could also couple to $D$ and $Z$ like $D^{c\prime}$. However, we have only one $D$, and so two of the mass eigenstates from $\{D^{c\prime},d_1^c,d_2^c\}$ remain light. We redefine them as the d-type quarks appearing in the MSSM. 

Were it not for the U(1)$_R$ symmetry, the following terms were admitted in the superpotential:
\dis{ \label{dim5}
u_i^cd_j^cD^{c\prime}
  ~,~~ 
Du_3^ce_3^c  ~~;~~~~q_iq_jD^{\prime} ~,~~ D^cq_3l_3 \qquad\qquad~~~ (i,j=1,2) ,
}
which are deduced to the operators leading to the dimension 5 proton decay,  $u_i^cu_3^cd_j^ce_3^c$ and $q_iq_jq_3l_3$ after integrating out $\{D, D^{c\prime}\}$ and $\{D^\prime, D^c\}$ of the $y_{D1}$ and $y_{D2}$ terms in Eq.~(\ref{W12}).\footnote{Only with the U(1)$_Z$ [and U(1)$_X$] gauge interactions, 
the dimension 6 operators leading to proton decay are  not induced. In this model, the dimension 6 proton decay is possible only through gravity interactions, which is still safe at the moment.} Although U(1)$_R$ is broken to the $Z_2$ symmetry, these are still forbidden, because all the superfields appearing in Eq.~(\ref{dim5}), including $\{D,D^c\}$ and $\{D^\prime, D^{c\prime}\}$, carry only the odd parity.  

%

Although the SM fermions can get their masses through the Yukawa couplings in Eqs.~(\ref{W3}) and (\ref{W12}), the mixings between the third and the first two families of them, and also the CP phase in the SM quark sector are absent only with the fields discussed above. 
Thus, we need one more vector-like lepton pair $\{L,L^c\}$, which are also assumed to be relatively heavier than other MSSM matter fields. 
%
%
How a mass term of the type $\mu_LLL^c$ can be obtained will be explained later. The mixings between the third and the first two families in the mass matrices of the d-type quarks and the charged leptons 
can be generated from the following Yukawa interactions after integrating out $\{D,D^{c\prime}\}$, $\{L,L^c\}$ and $\{H_u,H_d\}$, because $\nu_3^c$ develops a VEV: 
\dis{ \label{Wmix}
W_{\rm mix}=\sum_{i=1,2}\left(y_{Dd}\nu_3^cd_3^cD + y^i_{q}D^{c\prime}H_dq_i
+ y_{l}\nu_3^cl_3L^c + y^i_{Le}Lh_de_i^c\right) ,
}
which fills the $(3,i)$ entries of the mass matrices of the d-type quarks ($\equiv [M_D]_{3,i}$), and the $(i,3)$ of the charged leptons ($\equiv [M_E]_{i,3}$). 
They are estimated as $(y_{Dd}y_q^i/y_{D1})\langle\tilde{\nu}_3^c\rangle/\mu_H$ and $y_ly^i_{Le}\langle\tilde{\nu}_3^c\rangle/\mu_L$, respectively. 
Even if $[M_D]_{i,3}$ and $[M_E]_{3,i}$ ($i=1,2$) are still zero, $M_D^{\dagger}M_D$ and $M_E^{\dagger}M_E$ are fully general Hermitian matrices. Accordingly, the unitary matrices $U^{(d)}_{L}$ and $U^{(e)}_{L}$, which diagonalize  $M_D^\dagger M_D$ and $M_E^\dagger M_E$, respectively, and so the CKM and PMNS matrices describing the quark and lepton's mixings are also fully general in this model.   



Due to the U(1)$_Z$ and U(1)$_R$ symmetries, the right handed neutrino $\nu_3^c$ can not obtain the Dirac and Majorana masses. Hence, the seesaw mechanism should be implemented only with the two heavy right handed neutrinos $\nu_i^c$ ($i=1,2$) in \eq{W12}. As shown in Ref.~\cite{2RHnu}, the seesaw mechanism and also the leptogenesis can still work even with two heavy right handed neutrinos. The mixings of the observed neutrinos can come from the charged lepton sector.   

Finally, let us present various ``$\mu$ terms'' in this model, $\mu_HH_uH_d$, $\mu_LLL^c$, $\mu_YYY^c$, and $\mu_ZZZ^c$, whose generations are associated with the  SUSY breaking effect in the K${\rm \ddot{a}}$hler potential \cite{GM}: 
\dis{ \label{Kahler}
K=\frac{X^\dagger}{M_P}\left(\kappa_HH_uH_d +\kappa_LLL^c +\kappa_YYY^c +\kappa_ZZZ^c 
\right) + {\rm h.c.} , 
}
in which $X$ is a SUSY breaking source: its F-term component ($\equiv F_X$), which carries the U(1)$_R$ charge of $-2$, is assumed to develop a VEV of order $m_{3/2}M_P$. 
Thus, the VEV of $F_X$ breaks the U(1)$_R$ symmetry to $Z_2$, which can be interpreted as the matter parity in the MSSM, 
since the ordinary matter [and also $\{D^{(\prime)},D^{c(\prime)}\}$, $\{L,L^c\}$] except $\nu_3^c$ carry the U(1)$_R$ charges of odd integers.  
On the other hand, the superfields $S$, $\{h_d,h_u\}$, $\nu_3^c$,  $\{Y,Y^c\}$, and $\{Z,Z^{c}\}$, whose scalar components can develop VEVs, carry U(1)$_R$ charges of even integers. 
Accordingly, the remaining $Z_2$ symmetry forbids the R-parity violating couplings including the operators leading to dimension 4 proton decay, and also can guarantee the existence of the LSP dark matter.  
Note that the $\mu$ parameters in $\mu_HH_uH_d$, $\mu_{L}LL^c$, and $\mu_YYY^c$, which are all generated from \eq{Kahler}, 
can be regarded as spurion fields carrying the $2$ charge of U(1)$_R$, since they originate from the VEV of $F_X^*$.  

As mentioned above, the ``A-term'' corresponding to the $\lambda$ term in \eq{W3} provides a tadpole of $S$ in the scalar potential, when $\{h_u,h_d\}$ develop VEVs, and it eventually leads to a non-zero VEV of $S$. 
Such an ``A-term'' is induced by $W\supset \lambda XSh_uh_d/M_P$.
Since U(1)$_R$ as well as U(1)$_Z$ are broken, e.g.  additional tadpole terms of $S$ could be potentially induced. 
Indeed, an additional tadpole of $S$ is  
generated in the scalar potential at one loop, 
$\sim (\lambda y_{h1}^*y_{h2}^*/16\pi^2)\mu_H
\widetilde{S}(\widetilde{Z}^*)^2+{\rm h.c.}$, in which the fermionic components of $\{h_u,h_d\}$ and $\{H_d,H_u\}$ circulate in the loop. 
Thus, small enough $y_{h1,2}$ ($\lesssim 0.1$) leave intact our previous discussion on the size of $\langle\widetilde{S}\rangle$ and $\langle\widetilde{Z}\rangle$. 
Similarly, e.g. the ``A-terms'' of the $y_{h1,2}$ terms in \eq{W12}, and the ``B$\mu$-term'' of the $\kappa_H$ term in \eq{Kahler} 
are generated from $W\supset y_{h1,2}X^\dagger ZH_{u,d}h_{d,u}/M_P$, and $K\supset \kappa_HX^\dagger XH_uH_d/M_P^2+{\rm h.c.}$, respectively. 

The presence of $\kappa_{Y,Z}$ terms in \eq{Kahler} can affect our earlier discussion on the VEVs of 
$\{\tilde{\nu}_3^c, \widetilde{Y}, \widetilde{Y}^c,\widetilde{Z},\widetilde{Z}^c\}$. Only if $|\kappa_{Y,Z}|\lesssim 1$, however, their VEVs determined without the $\kappa_{Y,Z}$  would be  just slightly modified. 
As mentioned earlier, we require that $\kappa_{H,L}$ 
is relatively large ($\sim 5$). 
Throughout this paper, we assume that all the soft parameters at low energies are not heavier than $\mu_{H,L}$. 
$\{H_u,H_d\}$ make contribute to the radiative correction to the masses of ordinary Higgs $\{h_u,h_d\}$ via the $y_{h1;2}$ terms in \eq{W12}. If they were too heavy, hence, they  
could radiatively destabilize the electroweak scale. 
However, the mixings between $\{H_u,H_d\}$ and $\{h_u,h_d\}$, i.e. $y_{h1;2}$ in \eq{W12} are small enough ($\lesssim 0.3$), and so 
the electroweak scale still remain radiatively stable even with relatively large  $\kappa_{H,L}$.


In this model, the four fermionic components of $\{Z, Z^c, S\}$ and the U(1)$_Z$ gaugino, $( Z,  Z^c,  S, \lambda_Z)$    
are mixed to each other via the mass matrix: 
\dis{
\left(\begin{array}{cccc} 
0 & \mu_Z & 0 & \sqrt{2}g_Zq_z \langle \widetilde Z\rangle \\
\mu_Z & 0 & 0 & -\sqrt{2}g_Zq_z \langle\widetilde Z^c\rangle  \\ 
0 & 0 & 0 & \sqrt{2}g_Zq_s \langle \widetilde S\rangle\\
\sqrt{2}g_Zq_z \langle \widetilde Z\rangle  & -\sqrt{2}g_Zq_z\langle \widetilde Z^c\rangle &\sqrt{2}g_Z q_s \langle \widetilde S\rangle &  M_\lambda
\end{array}\right) ,
}  
where $q_z=2$, $q_s=4$, and $M_{\lambda}$ denotes the $U(1)_Z$ gaugino mass.
Here, we ignored $\lambda Sh_uh_d$ and $y_V\nu_3^cYZ^c$ couplings for a moment. 
For the case of $\langle \widetilde S\rangle \ll \langle \widetilde Z\rangle\sim \langle\widetilde Z^c\rangle\sim \mu_Z$, 
the mass of $S$-like (singlino-like) fermion 
is approximately given by  
\bea\label{Singlinomass1}
M_{ S} \approx  \frac{ 2 g_Z^2 q_s^2 \langle \widetilde S\rangle^2 }{
M_{\lambda} + 4g_Z^2 q_z^2\langle \widetilde{Z} \widetilde Z^c\rangle/\mu_Z} 
\sim \frac{\langle \widetilde S\rangle^2 }{\langle \widetilde Z\rangle} \sim \mu_{\rm eff}\left(\frac{\mu_{\rm eff} }{M_{Z}}\right) ,
\eea
where $M_{Z}$ stands for the U(1)$_Z$ symmetry breaking scale, and $\mu_{\rm eff}$ the 
effective $\mu$ parameter ($=\lambda\langle  \widetilde S\rangle$). The order of magnitude of Eq.~(\ref{Singlinomass1}) would still be the same, even if we include also the fermionic components of $\{\nu_3^c, Y, Y^c\}$ in the mass matrix.
Thus, we get $M_{S}\sim 100$ GeV for a
relatively large $\mu_{\rm eff}$ e.g. $\mu_{\rm eff}\sim 1$ TeV and $M_{Z}\sim 10$ TeV.
As a result, the invisible decay of the Higgs to the two singlinos is kinematically forbidden.  

On the other hand, if $\mu_{\rm eff}={\cal O}(100)$ GeV, the singlino's mass given by Eq.~(\ref{Singlinomass1}) becomes lighter than 10 GeV. In this case, the singlino mass would be dominantly given by the superpotential term $\lambda Sh_uh_d$: 
\bea\label{Singlinomass2}
M_{\tilde S}
\approx \frac{\lambda^2v_h^2{\rm sin}2\beta}{\mu_{\rm eff}+\lambda^2v_h^2/\mu_{\rm eff}} ~.
\eea
The upper bound of Eq.~(\ref{Singlinomass2}) is achieved when $\mu_{\rm eff}\sim \lambda v_h$. Thus,    
$M_{\tilde S} \lesssim  (\lambda v_h \sin2\beta)/2$. 
In the ``minimal" NMSSM case (without the U(1)$_Z$ gauge symmetry and extra matter), 
the perturbativity of $\lambda$ up to the GUT scale ($<0.7$) constrains the mass of the light singlino as $M_{\tilde S} < m_h/2\approx 63$ GeV. It opens the invisible decay of the Higgs to the two  singlinos \cite{PQNMSSM2}. 
In our model, however, the upper bound of $\lambda$ is relaxed, and so 
the singlino mass can be heavier than 63 GeV, maintaining the perturbativity of $\lambda$, 
e.g. $\lambda\approx 0.88$ for $n_Z=4$, ${\rm tan}\beta\approx 2$, and $\mu_{\rm eff}\approx 150$ GeV. Thus, the invisible decay of the Higgs can still be kinematically forbidden even for $\mu_{\rm eff}={\cal O}(100)$ GeV.


%

\underline{\bf Model of U(1)$_Z\times$U(1)$_X$} $\quad$ 
For the case that U(1)$_X$ is also considered, the first and second families of the MSSM chiral matter are still assumed to be neutral under U(1)$_X$ as well as    U(1)$_Z$, carrying unit U(1)$_R$ charges. 
We need more fields in the U(1)$_Z\times$U(1)$_X$ case, $\{Z^\prime,Z^{c\prime}\}$ and $\{N,N^c\}$. 
The charge assignments for the extra fields are displayed in Table \ref{tab:Qnumb3}.    
\begin{table}[!h]
\begin{center}
\begin{tabular}
{c|ccccccccc|cccc} 
{\rm Superfields}  &  $X$  &  $Y$  &  $Y^c$  &   $Z$  &
 $Z^c$  &  $D^{\prime}$   &  $D^{c\prime}$  &  $L$  &  $L^c$  &  $Z^\prime$  &  $Z^{c\prime}$  &
 $N$  &  $N^c$ 
  \\
\hline 
U(1)$_Z$ & ~$0$ & ~$1$ & $-1$ & ~$2$ & $-2$ & ~$0$ & ~$0$
 & ~$2$ & $-2$ & ~$2$ & $-2$ & $-1$ & ~$1$  \\
U(1)$_X$ & ~$0$ & $-3$ & ~$3$ & $-2$ & ~$2$ & ~$0$ & ~$0$
 & $-2$ & ~$2$ & ~$2$ & $-2$ & $-1$ & ~$1$  \\ 
U(1)$_{R}$ & ~$0$ & ~$0$ & ~$0$ & ~$0$ & ~$0$ & ~$3$ & ~$1$ & $-1$ & ~$1$ & ~$0$ & ~$0$ & $-2$ & ~$2$ 
\\
\end{tabular}
\end{center}\caption{
The charge assignments for the extra vector-like fields 
under the gauge U(1)$_Z\times$U(1)$_X$ and the global U(1)$_R$ in Case II. 
Here the U(1)$_Z$ and U(1)$_X$ charge normalizations, $n_Z/\sqrt{24}$ and $n_X/\sqrt{40}$ are omitted for simple presentations. 
The extra vector-like fields listed in Table \ref{tab:Qnumb2} should carry also U(1)$_X$ charges. In this case, more fields $\{Z^\prime,Z^{c\prime}\}$ and $\{N,N^c\}$ are needed.  As in Case I, however, the first two families of the chiral matter do not carry charges of U(1)$_Z\times$U(1)$_X$ but do unit charge of U(1)$_R$. 
}\label{tab:Qnumb3}
\end{table}
With the field contents in Table \ref{tab:Qnumb1} and \ref{tab:Qnumb3}, the beta function coefficients of U(1)$_Z$ and U(1)$_X$ are estimated as $b_Z=(4+\frac{1}{2})n_Z^2$ and $b_X=(4+\frac{3}{10})n_X^2$, respectively. 
They were utilized for analyses of the $v=2$ cases in Fig. 1-(b). 

For the U(1)$_Z\times$U(1)$_X$ case, the $y_{h2}$,   $y_{D1}$, and $y_V$ terms in Eq.~(\ref{W12}), and $y_l^i$ term in Eq.~(\ref{Wmix}) should be modified with the new superfields: 
\dis{
&y_{h2}ZH_dh_u \rightarrow y_{h2}Z^\prime H_dh_u ~,\quad y_{D1}ZDD^{c\prime} \rightarrow  
y_{D1}Z^\prime DD^{c\prime} ~, 
\\
&~~ y_{V}\nu_3^cYZ^c \rightarrow y_{V}\nu_3^cYZ^{c\prime} ~, 
\qquad ~  y^i_{l}\nu_3^cl_3L^c \rightarrow y^i_{l}N^cl_3L^c . 
}
Instead of $Z^c$, thus, $Z^{c\prime}$ obtains a VEV together with $\nu_3^c$ and $Y$ from the ``A-term'' of $y_V$. From the D-term potentials, $Z^\prime$ and $Y^c$ can also get VEVs.  
With the new superfields, the superpotential allows the following term:
\dis{
W\supset y_NYZ^cN^c ,
}
whose ``A-term'' can induce VEVs of $Z^c$ and $N^c$ as well as $Y$. Then, the D-term potentials can yield VEVs of $Z$ and $N$. 
The K${\rm \ddot{a}}$hler potential Eq.~(\ref{Kahler}) is supplemented with  
\dis{
K\supset \frac{X^\dagger}{M_P}\left(\kappa_{Z^\prime}Z^\prime Z^{c\prime}
+\kappa_NNN^c\right)  + {\rm h.c.} , 
}
which generates the ``$\mu$ terms'' for $\{Z^\prime,  Z^{c\prime}\}$ and $\{N,N^c\}$.

\section{Breaking scale of extra U(1)} \label{sec:FCNC}

Now let us discuss the U(1)$_Z$ breaking scale. 
Due to the family dependent charge assignment of U(1)$_Z$,
the  flavor violating process can be induced through 
the exchange of the U(1)$_Z$ gauge boson $Z_\mu'$.
We will estimate $\langle \tilde Z^c\rangle$, which determines the U(1)$_Z$ breaking scale, 
following the the formulation driven by Ref.~\cite{Langacker:2000ju}. 
A similar estimation would be applicable to the model of U(1)$_Z\times$U(1)$_X$.

The most important constraint on the mixing angle $\theta_Z$ between $Z_\mu$ and $Z'_\mu$ vector bosons 
come from the coherent $\mu$-$e$ conversion rate in nuclei. 
The branching ratio is \cite{Bernabeu:1993ta,Kuno:1999jp}
\bea
&&\qquad~~ {\rm Br}(\mu {\cal N}\to e{\cal N}^*)=\frac{\alpha_{\rm em}^3|F_p|^2}{2\pi^2} \frac{G_F^2 m_\mu^5}{\Gamma_{\rm cap}}
\frac{Z_{\rm eff}^4}{Z_{\rm atm}}
\Big(|Q^{e_L}_{12}|^2 + |Q^{e_R}_{12}|^2 \Big) \\
&&\times\left| x_1\Big(\frac{Z_{\rm atm}-N_{\rm ncl}}{2} - 2 Z_{\rm atm} \sin^2\theta_W\Big) 
+ x_2 (Z_{\rm atm}+ 2N_{\rm ncl})\Big(Q^{d_L}_{11} + Q^{d_R}_{11}\Big) \right|^2,
\nonumber 
\eea
where $F_p$ is the nuclear form factor, 
$\Gamma_{\rm cap}$ is the muon capture rate, $Z_{\rm eff}$ is the 
effective chage of the nuclei for the muon, $Z_{\rm atm}$ ($N_{\rm ncl}$) is the atomic number (the neutron number) 
of the nucleus ${\cal N}$. 	
$Q^{d_L}_Z $ and $Q^{e_L}_Z $ are defined with the CKM and PMNS matrices, 
$(Q^{d_L}_Z)_{ij}\equiv q_m (V_{\rm CKM})^*_{3i}(V_{\rm CKM})_{3j}$ and $(Q^{e_L}_Z)_{ij} \equiv q_m (V_{\rm PMNS})_{i3}(V_{\rm PMNS})^*_{j3}$, 
where $q_m$ denotes the electromagnetic charge of relevant SM matter. 
On the other hand, $(Q^{d_R}_Z)_{ij}$ and $(Q^{e_R}_Z)_{ij}$ are rather model dependent. 
We assume that $(Q^{d_R}_Z)_{ij} $ and $(Q^{e_R}_Z)_{ij}$
are the same order with $(Q^{d_L}_Z)_{ij}$ and  $(Q^{e_L}_Z)_{ij}$, respectively. 
For a small mixing ($\theta_Z\ll 1$), $x_1$ and $x_2$ are given by
\bea
x_1\approx \frac{g_Z  \theta_Z}{\sqrt{g^2 +g'^2}},\quad
x_2\approx \frac{v_h^2}{2 (M_{Z'}^2/g_Z^2)},
\eea 
where $M_{Z'}$ denotes the mass of the U(1)$_Z$ gauge boson.  
From the SINDRUM II collaboration, we get the most serious bound from ${\cal N}$=Au \cite{Bertl:2006up}:
\dis{
{\rm Br}(\mu{\rm Au}\to e{\rm Au}) < 7 \times 10^{-13} ,
}
which yields the constraint on $x_1$ as 
\dis{ \label{sundrum-II}
x_1 \lesssim 2\times 10^{-6} .
}
If the kinetic mixing between U(1)$_Y$ and U(1)$_Z$ is negligible, $x_1\approx 0$ at tree-level, since $\widetilde{Z}^c$ does not carry the SM gauge quantum number. Thus, it can be easily satisfied in this model. 

The constraint on $x_2$ from the quark sector is rather mild. 
Since the U(1)$_Z$ charges for the first and second families of the SM chiral fields are zero, 
the flavor violating effects are proportional to the off-diagonal components of the CKM matrix.  
However, there is no such large suppression in the lepton sector.  
The most important constraint originates from the 
flavor changing muon decay modes: $\mu^-\to e^-\gamma$ 
and $\mu^-\to e^-e^+e^-$.
The branching ratios for the processes are estimated as 
\bea
{\rm Br}(\mu^-\rightarrow e^-\gamma) &\approx& \frac{48 \alpha_{\rm em}}{\pi} (q_m^2x_2)^2 
 \left|(V_R^e)_{31} (V_{\rm PMNS})_{23}\frac{[M_E]_{3,3}}{m_\mu}\right|^2 ,
\nonumber \\
{\rm Br}(\mu^-\rightarrow e^-e^+e^-)&\approx&4 x_2^2\left\{ 2\Big|(Q^{e_L}_Z)_{11}( Q^{e_L}_Z)_{12}\Big|^2
+2\Big|(Q^{e_R}_Z)_{11}(Q^{e_R}_Z)_{12}\Big|^2\right. \\
&&\left.
+\, \Big|(Q^{e_L}_Z)_{11}(Q^{e_R}_Z)_{12}\Big|^2
+\Big|(Q^{e_R}_Z)_{11}(Q^{e_L}_Z)_{12}\Big|^2 \right\}  \nonumber 
\eea
in the limit of a small $Z_\mu$-$Z_\mu'$ mixing, which is already constrained by \eq{sundrum-II}. 
Here $V_R^e$ indicates the unitary matrix diagonalizing $M_EM_E^\dagger$ (rather than $M_E^\dagger M_E$), which does not 
contribute to the PMNS matrix.  
$[M_E]_{3,3}$ means the $(3,3)$ components in the mass matrix of the charged leptons ($\equiv y_\tau v_h{\rm cos}\beta$).
The present experimental bounds for such processes \cite{Adam:2013mnn, Bellgardt:1987du} are 
found in Table \ref{tab:bound},
which provide the most stringent constraints on $x_2$ 
or $\langle \widetilde{Z}^c\rangle$:
\dis{
\langle \widetilde{Z}^c\rangle \gtrsim 5.6~{\rm TeV}\times \left(\frac{|y_\tau|}{10^{-3}}\right)^{1/2}\left(\frac{{\rm cos}\beta}{0.2}\right)^{1/2} , 
}  
and $\langle \widetilde{Z}^c\rangle \gtrsim 3.1~{\rm TeV}$, respectively.  
Thus, the VEV of $\widetilde{Z}^c$ discussed in section \ref{sec:model} ($\approx 5-10~{\rm TeV}$) 
is consistent with these bounds. 
Other less severe constraints on $x_2$ \cite{Beringer:1900zz} are also listed in Table \ref{tab:bound}.  

\begin{table}[!h]
\begin{center}
\begin{tabular}
{c|c} 
\hline
Experimental bounds  &  ~$q_m^2x_2$ ($=v_h^2/16\langle \widetilde{Z}^c\rangle^2$)
\\ \hline 
${\rm Br}(\mu^-\rightarrow e^-\gamma) < 5.7 \times 10^{-13}$  & $~< 0.2\, \frac{m_\mu}{|[M_E]_{3,3}|} \times 10^{-4}$ 
\\
${\rm Br}(\mu^-\rightarrow e^-e^+e^-) < 1.0\times 10^{-12}$  & $< 2 \times 10^{-4}$
\\
\hline
$\Delta m_{B^0_s} =  (117.0 \pm 0.8)
 \times 10^{-13}\, {\rm GeV}$ & $< 10^{-3}$
\\
$\Delta m_{B^0}  =  (3.337\pm 0.033)
 \times 10^{-13}\, {\rm GeV}$ & $< 10^{-3}$
\\
$\Delta m_{K^0}= (3.484\pm 0.006)\times 10^{-15}\, {\rm GeV}$ & $< 10^{-2}$
\\
\hline
$\epsilon_K = (2.233\pm 0.015)\times 10^{-3}$ & $< 4\times 10^{-4}$
\\
\hline 
${\rm Br}(B^0\to \mu^+\mu^-) < 1.4\times 10^{-9}$ & $< 10^{-2}$ 
\\
${\rm Br}(K^0_L\to\mu^\pm e^\mp) < 4.7\times 10^{-12}$ & $< 1.5 \times 10^{-3}$
\\
${\rm Br}(K^0_L\to e^+e^- \pi^0) < 2.8\times 10^{-10}$ &  $< 0.4$
\\
${\rm Br}(K^0_L \to \mu^+\mu^-\pi^0)< 3.8\times 10^{-10}$ & $< 0.1$
\\
${\rm Br}(K^0_L\to \mu^+e^-\pi^0) < 7.6\times 10^{-11}$ & $< 0.1$

\\
\hline
\end{tabular}
\end{center}\caption{Constraints on  $\langle \widetilde{Z}^c\rangle$ from various experimental bounds. $m_\mu$ denotes the muon mass, and $[M_E]_{3,3}$ means the $(3,3)$ component in the mass matrix of the charged leptons ($\equiv y_\tau v_h{\rm cos}\beta$). 
}\label{tab:bound}
\end{table}
The upper bound on $x_2$ will possibly be further  lowered by future experiments. Then, this model will be testable. 
Especially for $\mu\to 3e$, the sensitivity 
is planned to reach ${\rm Br}(\mu\to 3e)=10^{-16}$ \cite{Mu3e}. 
Concerning the $\mu$-$e$ conversion, many experiments are proposed to explore 
${\rm Br}(\mu{\rm Al}\to e{\rm Al}) = 10^{-16}$ \cite{comet,mu2e} and 
${\rm Br}(\mu{\rm Ti}\to e{\rm Ti})=10^{-18}$ \cite{Kuno:2013mha}.

Another potential flavor violation effects in this model are the processes mediated by the heavy Higgs $\{H_u,H_d\}$. 
The constraints by them would be mild compared to those by $Z'$ boson mediations, 
because small Yukawa couplings are also involved there. 
%
%
In the mass eigen basis, the neutral flavor-changing couplings can be written as 
\bea \label{fcnc}
W_{FCNC} =\sum_{i,j=1,2,3}\left(\xi^d_{ij}\hat{d}_{i}\hat  {d}_{j}^c
+\xi^e_{ij}\hat{e}_{i}\hat{e}_{j}^c \right)H_d^0 , 
\eea
where $\xi^{d,e}_{ij}$ parametrize (Yukawa couplings)$\times$(mixing angles), and so they are of order the Yukawa couplings or smaller. 
Here $H_d^0$ denotes the neutral component of $H_d$.
After integrating out the heavy Higgs $\{H_u,H_d\}$,  
\eq{fcnc} provides flavor violating four fermion interactions 
suppressed by $\mu_H^2$ from the K${\rm\ddot{a}}$hler potential.   
Unlike $H_d^0$, $H_u^0$ does not give rise to flavor violations. 
Note that $H_u^0$ has only the $2\times 2$ block-diagonal couplings
of the first and second generations of the SM chiral matter, while $h_u^0$ couples only to the third generation. 

Due to the hierarchical structure of the masses for the quarks and charged leptons, 
we employ the parametrization of Cheng and Sher  \cite{Cheng:1987rs} as   
\bea 
\xi^{d}_{ij}=\lambda^{d}_{ij}\frac{\sqrt{m^d_i m^d_j}}{v_h} ,\quad
\xi^{e}_{ij}=\lambda^{e}_{ij}\frac{\sqrt{m^e_i m^e_j}}{v_h} ,
\eea
where $m^q_i$ ($m^e_i$) denotes the mass of d-type quark (charged lepton) of the generation $i$. 
For $\lambda^{d,e}\sim{\cal O}(1)$, thus, $\xi^{d,e}$ 
become the same order of magnitude with the average of the relevant Yukawa couplings. 
If $\lambda^{d,e}$ turn out to be of order unity, hence, the couplings in \eq{fcnc} can be regarded as being natural. 
On the other hand, if the experimental bounds requires  too small $\lambda^{d,e}$, the couplings in \eq{fcnc} 
should be finely tuned. 

For the quark sector, the most serious bound comes from the neutral meson ($M^0= q_i\bar q_j$) mixing. 
The mass splitting is calculated as \cite{Gupta:2009wn}
\bea
\frac{\Delta m_{M^0}}{m_{M^0}} \approx 
 \frac{(\xi_{ij}^d)^2 B_{M^0}F_{M^0}^2 }{3\mu_H^2  }
 \left[1+\frac{6 m_{M^0}^2}{(m^q_i+m^q_j)^2}\right] ,
\eea
where $F_{M^0}$ is the meson decay constant, and $B_{M^0}$ is the vacuum 
insertion parameters defined in \cite{Atwood:1996vj}. The experimental bounds on $\Delta m_{B^0_s}$, $\Delta m_{B^0}$ and $\Delta m_{K^0}$ in Table \ref{tab:bound}
constrain the parameters as 
\bea
(\lambda^d_{12},\   \lambda^d_{31},\ \lambda^d_{32})
\left(\frac{5\, {\rm TeV}}{\mu_H}\right) 
\lesssim (4.2,\   2.5,\  2.5) . 
\eea
The less severe constraint is from ${\rm Br}(B_s^0\to \mu^+\mu^-)< 6.4\times 10^{-9}$ \cite{Beringer:1900zz}, 
which yields 
\bea
\sqrt{\lambda^d_{32}\lambda^e_{22}}\left(\frac{5\,{\rm TeV}}{\mu_H}\right) \lesssim 37.
\eea
For the lepton sector, the most important constraint is from the $\mu\to e\gamma$ bound.
The decay rate is given by \cite{Diaz:2002uk}
\bea
\Gamma(\mu\to e\gamma) \approx \left[\frac{\alpha_{\rm em} (\xi^e_{13})^2(\xi^e_{23})^2}{4\pi^4}\right]
\left[\frac{m_\tau^4 m_\mu}{\mu_H^4}\right]\left[\ln\frac{\mu_H}{m_\tau}\right]^2 .
\eea
So the resulting bound is estimated as 
\bea
\sqrt{\lambda_{13}^e\lambda_{23}^e}\left(\frac{5\,{\rm TeV}}{\mu_H}\right) \lesssim 3.3 .
\eea
Therefore, all the flavor violations associated with the heavy Higgs, $\{H_u,H_d\}$ can be sufficiently suppressed,
only if $\mu_H\gtrsim 5~{\rm TeV}$.

\section{Conclusion} \label{sec:conclusion}

We considered the perturbative U(1) gauge extensions of the NMSSM to relieve the LP constraint on the ``$\lambda$'' coupling, maintaining the gauge coupling unification. 
They are closely associated with raising the tree-level mass of the Higgs. To minimize the beta function coefficient(s) of U(1)$_Z$ or U(1)$_Z\times$U(1)$_X$, which is necessary for U(1) gauge interaction(s) relevant down to low energies, 
we assign U(1) charges only to the Higgs doublets and the third family of the chiral matter among the MSSM field contents. 
In the U(1)$_Z$ [U(1)$_Z\times$U(1)$_X$] case, the low energy value of $\lambda$ can be lifted up to $0.85-0.95$ [$0.9-1.0$], depending on the normalization of the charges, when the gauge coupling $g_{Z}$ [and also $g_X$] and $\lambda$ are constrained not to blow up below the GUT scale. 
We also discussed how to induce the Yukawa couplings for the first and second families of the quarks and leptons, and the resulting phenomenological constraints associated with flavor violations.

\acknowledgments

This research was supported by Basic Science Research Program through the 
National Research Foundation of Korea (NRF) funded by the Ministry of Education, Grant No. 2013R1A1A2006904 (B.K.) and 2011-0011083 (C.S.S.), and also in part by Korea Institute for Advanced Study (KIAS) grant funded by the Korea government 
(B.K.).
C.S.S. acknowledges the Max Planck Society (MPG), the Korea Ministry of Education, Gyeongsangbuk-Do, and Pohang City for the support of the Independent Junior Research Group at the Asia Pacific Center for Theoretical Physics (APCTP).


\end{document}